\documentclass[12pt,a4paper]{article}
\usepackage{latexsym,amssymb,amsmath}

\begin{document}
\title{On the circuit complexity of the standard and the Karatsuba methods of multiplying integers\footnote{Translated
 from the Russian original published in: Proc. of XXII Conf. ``Information means and technology'' (Moscow, November 18--20, 2014). Vol. 3. Moscow, MPEI, 2014, 180--187.}}
\date{}
\author{Igor S. Sergeev\footnote{e-mail: isserg@gmail.com}}
\maketitle

The goal of the present paper is to obtain accurate estimates for
the complexity of two practical multiplication methods: standard
(school) and Karatsuba~\cite{ka}. Here, complexity is the minimal
possible number of gates in a logic circuit implementing the
required function over the basis $\{AND, OR, XOR, NAND, NOR,
XNOR\}$.

One can find upper estimates for the said methods e.g.
in~\cite{ga}. The standard method has complexity $M(n) \le
6n^2-8n$. In the case $n=2^k$, the complexity $K(n)$ of the
Karatsuba method can be deduced from the recursion $K(2n) \le
3K(n)+49n-8$ as $K(2^k) \le 26\frac29\cdot 3^k - 49\cdot2^k+4 $.

We intend to show that the above estimates may be improved with
the help of the result~\cite{dkky} stating that a sum of $n$ bits
may be computed via $4.5n$ operations instead of $5n$ as in the
naive approach. The resulting bounds are $M(n) \le
5.5n^2-6.5n-1+(n\bmod 2)$ and
\begin{equation}\label{kar}
K(2^k) \le 25\frac{83}{405}\cdot3^k - 38\cdot2^k -
\frac{81}5\Phi_{k+2} - \frac{37}5\Phi_{k+1} + 20,
\end{equation}
where $\{\Phi_k\}$ is the Fibonacci sequence: $\Phi_1=\Phi_2=1$,
$\Phi_{k+2}=\Phi_{k+1}+\Phi_k$.

{\bf Auxiliary circuits.} The circuits below are built from the
following subcircuits: half-adders $HA$, $HA^{\pm}$, (3,\,2)-carry
save adders (CSA) $FA_3$, $FA_3^-$, $FA_3^0$, $SFA_3$, $SFA_3^-$,
(5,\,3)-CSA $MDFA$, $MDFA^-$ (the $MDFA$ circuit was proposed
in~\cite{dkky}). Specifically, they implement the functions:

$HA$: $(x_1,\,x_2) \to (u;\,v)$, where $x_1+x_2 = 2u+v$;

$HA^{\pm}$: $(x_1,\,x_2) \to (u;\,v)$, where $x_1-x_2 = -2u+v$;

$FA_3$: $(x_1,\,x_2,\,x_3) \to (u;\,v)$, where $x_1+x_2+x_3 =
2u+v$;

$FA_3^-$: $(x_1,\,x_2,\,x_3) \to (u;\,v)$, where $x_1+x_2-x_3 =
2u-v$;

$FA_3^0$: $(x_1,\,x_2,\,x_3) \to (u;\,v)$, where $x_1+x_2-x_3 =
2u+v$, if $x_1+x_2-x_3 \ge 0$;

$SFA_3$: $(x_1,\,x_1\oplus x_2,\,x_3) \to (u;\,v)$, where
$x_1+x_2+x_3 = 2u+v$;

$SFA_3^-$: $(x_1,\,x_1\oplus x_2,\,x_3) \to (u;\,v)$, where
$x_1-x_2+x_3 = 2u-v$;

$MDFA$: $(x_1,\,x_1\oplus y_1,\,x_2,\,x_2\oplus y_2,\,z) \to
(u_1,\,u_1\oplus u_2;\,v)$, where $x_1+y_1+x_2+y_2+z =
2(u_1+u_2)+v$;

$MDFA^-$: $(x_1,\,x_1\oplus y_1,\,x_2,\,x_2\oplus y_2,\,z) \to
(u_1,\,u_1\oplus u_2;\,v)$, where $x_1-y_1+x_2-y_2+z =
2(u_1-u_2)+v$.

These circuits are shown on Fig. 1. Gates $AND, OR, XOR$ are
denoted by symbols $\wedge$, $\vee$, $\oplus$, respectively.
Inverted inputs are marked by small circles.

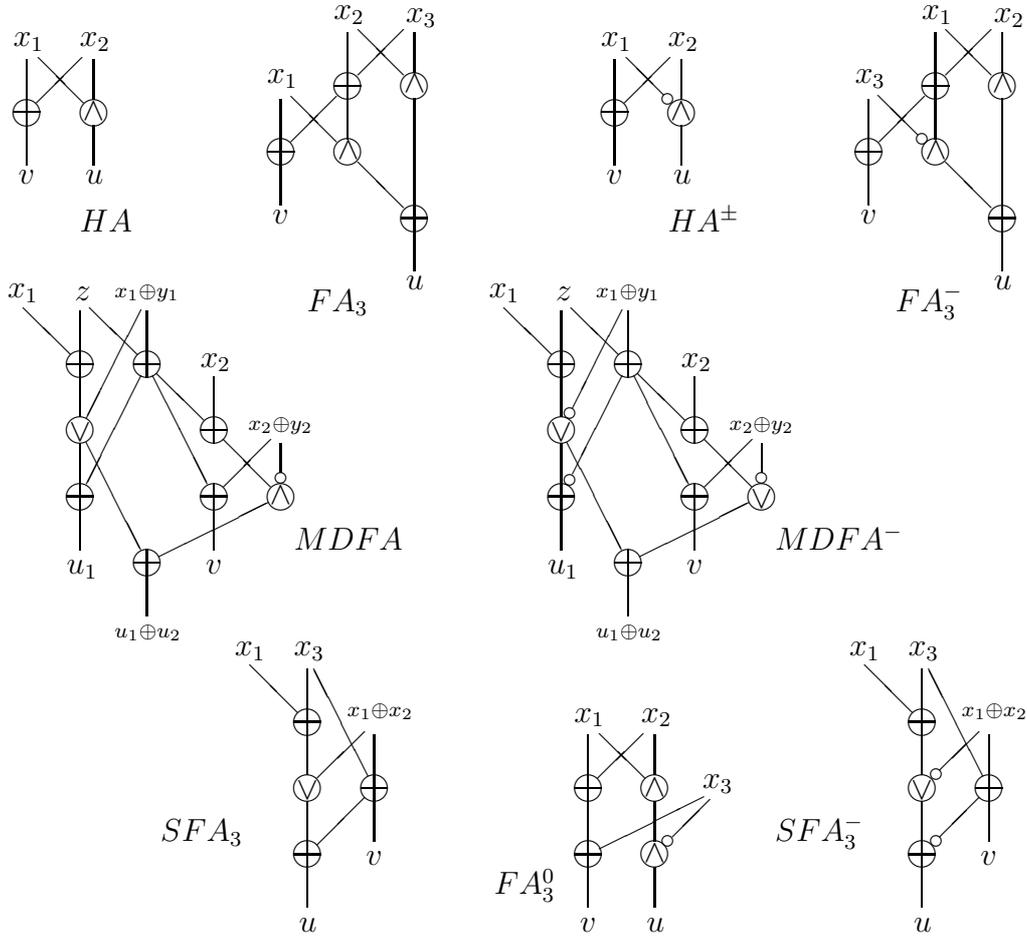
\begin{figure}[htb]
\begin{picture}(400,350)(0,-40)

\put(0,240){\begin{picture}(50,60)
\multiput(10,25)(25,0){2}{\circle{10}}
\multiput(10,30)(25,0){2}{\line(0,1){15}}
\multiput(10,5)(25,0){2}{\line(0,1){15}}
\put(13.7,28.7){\line(1,1){17}} \put(31.3,28.7){\line(-1,1){17}}
\put(5,25){\line(1,0){10}} \put(10,20){\line(0,1){10}}
\put(5,50){$x_1$} \put(30,50){$x_2$} \put(7,-2){$v$}
\put(32,-2){$u$} \put(31,22.2){$\wedge$}
\end{picture}}

\put(30,220){$HA$}

\put(220,240){\begin{picture}(50,60)
\multiput(10,25)(25,0){2}{\circle{10}}
\multiput(10,30)(25,0){2}{\line(0,1){15}}
\multiput(10,5)(25,0){2}{\line(0,1){15}}
\put(13.7,28.7){\line(1,1){17}} \put(28.3,31.7){\line(-1,1){14}}
\put(29.8,30.2){\circle{4}} \put(5,25){\line(1,0){10}}
\put(10,20){\line(0,1){10}} \put(5,50){$x_1$} \put(30,50){$x_2$}
\put(7,-2){$v$} \put(32,-2){$u$} \put(31,22.2){$\wedge$}
\end{picture}}

\put(250,220){$HA^{\pm}$}

\put(80,200){\begin{picture}(100,100)
\multiput(50,75)(25,0){2}{\circle{10}}
\multiput(25,50)(25,0){2}{\circle{10}} \put(75,25){\circle{10}}
\multiput(25,30)(25,25){3}{\line(0,1){15}}
\multiput(25,55)(25,25){2}{\line(0,1){15}}
\multiput(29,54)(25,25){2}{\line(1,1){17}}
\multiput(71,29)(-25,25){2}{\line(-1,1){17}}
\put(71,79){\line(-1,1){17}} \put(75,70){\line(0,-1){40}}
\put(75,20){\line(0,-1){15}}
\multiput(20,50)(25,25){2}{\line(1,0){10}}
\put(70,25){\line(1,0){10}} \put(75,20){\line(0,1){10}}
\multiput(25,45)(25,25){2}{\line(0,1){10}}
\multiput(46.3,47.5)(25,25){2}{$\wedge$} \put(20,76){$x_1$}
\put(45,100){$x_2$} \put(72,99){$x_3$} \put(22,23){$v$}
\put(72,-2){$u$} \end{picture}}

\put(115,190){$FA_3$}

\put(300,200){\begin{picture}(100,100)
\multiput(50,75)(25,0){2}{\circle{10}}
\multiput(25,50)(25,0){2}{\circle{10}} \put(75,25){\circle{10}}
\multiput(25,30)(25,25){3}{\line(0,1){15}}
\multiput(25,55)(25,25){2}{\line(0,1){15}}
\multiput(29,54)(25,25){2}{\line(1,1){17}}
\put(43.5,56.5){\line(-1,1){15}}
\put(45,55){\circle{4}}
\put(71,29){\line(-1,1){17}}
\put(71,79){\line(-1,1){17}} \put(75,70){\line(0,-1){40}}
\put(75,20){\line(0,-1){15}}
\multiput(20,50)(25,25){2}{\line(1,0){10}}
\put(70,25){\line(1,0){10}} \put(75,20){\line(0,1){10}}
\multiput(25,45)(25,25){2}{\line(0,1){10}}
\multiput(46.3,47.5)(25,25){2}{$\wedge$} \put(20,76){$x_3$}
\put(45,100){$x_1$} \put(72,99){$x_2$} \put(22,23){$v$}
\put(72,-2){$u$} \end{picture}}

\put(335,190){$FA_3^-$}

\put(0,70){\begin{picture}(110,130)
\multiput(30,50)(0,25){3}{\circle{10}}
\multiput(55,100)(25,-25){3}{\circle{10}}
\multiput(80,50)(-25,-25){2}{\circle{10}}
\multiput(30,30)(0,25){4}{\line(0,1){15}}
\multiput(55,5)(0,100){2}{\line(0,1){15}}
\multiput(80,30)(0,50){2}{\line(0,1){15}} \put(105,57){\circle{4}}
\put(105,59){\line(0,1){11}}
\multiput(101.3,53.7)(-25,25){3}{\line(-1,1){17.7}}
\put(26.3,103.7){\line(-1,1){17.7}}
\multiput(32.2,70.6)(25,25){2}{\line(1,-2){20.4}}
\multiput(52.8,95.6)(0,25){2}{\line(-1,-2){20.4}}
\put(83.7,53.7){\line(1,1){17.7}}
\put(59.4,27.2){\line(2,1){40.9}} \put(3,125){$x_1$}
\put(28,124){$z$} \put(43,125){$\scriptstyle x_1\oplus y_1$}
\put(75,99){$x_2$} \put(93,75){$\scriptstyle x_2\oplus y_2$}
\multiput(30,45)(0,50){2}{\line(0,1){10}}
\multiput(25,50)(0,50){2}{\line(1,0){10}}
\multiput(55,20)(0,75){2}{\line(0,1){10}}
\multiput(50,25)(0,75){2}{\line(1,0){10}}
\multiput(80,45)(0,25){2}{\line(0,1){10}}
\multiput(75,50)(0,25){2}{\line(1,0){10}} \put(101,47.2){$\wedge$}
\put(26,70.9){$\vee$} \put(77,21){$v$} \put(25,21){$u_1$}
\put(43,-3){$\scriptstyle u_1\oplus u_2$}
\end{picture}}

\put(110,100){$MDFA$}

\put(180,70){\begin{picture}(110,130)
\multiput(30,50)(0,25){3}{\circle{10}}
\multiput(55,100)(25,-25){3}{\circle{10}}
\multiput(80,50)(-25,-25){2}{\circle{10}}
\multiput(30,30)(0,25){4}{\line(0,1){15}}
\multiput(55,5)(0,100){2}{\line(0,1){15}}
\multiput(80,30)(0,50){2}{\line(0,1){15}} \put(105,57){\circle{4}}
\put(105,59){\line(0,1){11}}
\multiput(101.3,53.7)(-25,25){3}{\line(-1,1){17.7}}
\put(26.3,103.7){\line(-1,1){17.7}}
\multiput(32.2,70.6)(25,25){2}{\line(1,-2){20.4}}
\multiput(52.8,95.6)(0,25){2}{\line(-1,-2){18.6}}
\multiput(33.2,81.4)(0,-25){2}{\circle{4}}
\put(83.7,53.7){\line(1,1){17.7}}
\put(59.4,27.2){\line(2,1){40.9}} \put(3,125){$x_1$}
\put(28,124){$z$} \put(43,125){$\scriptstyle x_1\oplus y_1$}
\put(75,99){$x_2$} \put(93,75){$\scriptstyle x_2\oplus y_2$}
\multiput(30,45)(0,50){2}{\line(0,1){10}}
\multiput(25,50)(0,50){2}{\line(1,0){10}}
\multiput(55,20)(0,75){2}{\line(0,1){10}}
\multiput(50,25)(0,75){2}{\line(1,0){10}}
\multiput(80,45)(0,25){2}{\line(0,1){10}}
\multiput(75,50)(0,25){2}{\line(1,0){10}} \put(101,45.9){$\vee$}
\put(26,70.9){$\vee$} \put(77,21){$v$} \put(25,21){$u_1$}
\put(43,-3){$\scriptstyle u_1\oplus u_2$}
\end{picture}}

\put(290,100){$MDFA^-$}

\put(210,-40){\begin{picture}(70,80)
\multiput(10,25)(0,25){2}{\circle{10}}
\multiput(35,25)(0,25){2}{\circle{10}}
\multiput(10,20)(0,25){2}{\line(0,1){10}}
\multiput(5,25)(0,25){2}{\line(1,0){10}}
\multiput(31,22.2)(0,25){2}{$\wedge$}
\multiput(10,5)(0,25){3}{\line(0,1){15}}
\multiput(35,5)(0,25){3}{\line(0,1){15}}
\put(13.7,53.7){\line(1,1){17}} \put(31.3,53.7){\line(-1,1){17}}
\put(39.8,30.2){\circle{4}} \put(41.7,31.7){\line(1,1){14}}
\put(14.4,27.2){\line(2,1){39}} \put(5,75){$x_1$}
\put(30,75){$x_2$} \put(53,50){$x_3$} \put(7,-4){$v$}
\put(32,-4){$u$}
\end{picture}}

\put(185,-30){$FA_3^0$}

\put(80,-40){\begin{picture}(80,110)
\multiput(35,50)(0,25){2}{\circle{10}}
\multiput(35,25)(25,25){2}{\circle{10}}
\multiput(35,5)(0,25){4}{\line(0,1){15}}
\multiput(60,30)(0,25){2}{\line(0,1){15}}
\multiput(38.7,28.7)(0,25){2}{\line(1,1){17.7}}
\put(31.3,78.7){\line(-1,1){17.7}}
\put(57.8,54.4){\line(-1,2){20.4}}
\multiput(35,20)(0,50){2}{\line(0,1){10}}
\multiput(30,25)(0,50){2}{\line(1,0){10}}
\put(55,50){\line(1,0){10}} \put(60,45){\line(0,1){10}}
\put(31,45.9){$\vee$} \put(8,100){$x_1$} \put(30,100){$x_3$}
\put(50,77){$\scriptstyle x_1\oplus x_2$} \put(57,21){$v$}
\put(32,-4){$u$}
\end{picture}}

\put(60,-10){$SFA_3$}

\put(310,-40){\begin{picture}(80,110)
\multiput(35,50)(0,25){2}{\circle{10}}
\multiput(35,25)(25,25){2}{\circle{10}}
\multiput(35,5)(0,25){4}{\line(0,1){15}}
\multiput(60,30)(0,25){2}{\line(0,1){15}}
\multiput(41.7,31.7)(0,25){2}{\line(1,1){14.7}}
\put(31.3,78.7){\line(-1,1){17.7}}
\put(57.8,54.4){\line(-1,2){20.4}}
\multiput(35,20)(0,50){2}{\line(0,1){10}}
\multiput(30,25)(0,50){2}{\line(1,0){10}}
\put(55,50){\line(1,0){10}} \put(60,45){\line(0,1){10}}
\put(31,45.9){$\vee$} \put(8,100){$x_1$} \put(30,100){$x_3$}
\put(50,77){$\scriptstyle x_1\oplus x_2$} \put(57,21){$v$}
\put(32,-4){$u$} \multiput(40.2,30.2)(0,25){2}{\circle{4}}
\end{picture}}

\put(290,-10){$SFA_3^-$}

\end{picture}
\caption{Auxiliary circuits}
\end{figure}

{\bf Standard method.} The first stage of the standard method of
multipli\-ca\-tion of $n$-bit integers involves $n^2$ bit
multiplications. The next stage~--- multiple addi\-tion~--- may be
performed via summation of bits in consecutive columns (column is
a set of bits of the same order). The summation utilizes the
aforementioned auxiliary subcircuits. The result of a column
summation is a bit of the product and a set of carries to the next
order.

Let us index columns from 1 in increasing order. Then, after the
first phase one has $n-|n-k|$ bits in a $k$-th column,
$k=1,\ldots,2n-1$. Consider the following rule of column
summation: if there exist 5 summand bits, use $MDFA$; else, if
there exist 3 summand bits, use $FA_3$; else, if there exist 2
summand bits, use $HA$.

Denote by $h(k)$ number of summand bits in the $k$-th column after
completion of summation in all lower-order columns. Clearly,
$h(1)=1$. Let us check by induction that $h(k)=2k-2$ for $2 \le k
\le n$. Obviously, the statement holds for $k=2$. Assume, it also
holds for $k=t$ and consider summation in the $t$-th column. By
the declared strategy, summation of $2t-2$ bits involves $\lfloor
t/2 \rfloor -1$ circuits $MDFA$, one circuit $HA$, and in the case
of odd $t$, one more circuit $FA_3$. In total, it produces $t-1$
carries to the next order. Hence, $h(t+1)=t+1+t-1=2(t+1)-2$, as
required.

By analogy, we conclude that $h(n+1)=2n-2$ and $h(2n-k) = 2k+1$
for $0\le k \le n-2$. For summation in the $(n+1)$-th column one
use the same set of circuits as for the $n$-th column. For
summation in $(2n-k)$-th column we use $\lfloor k/2 \rfloor$
circuits $MDFA$, and in the case of odd $k\ne1$ an additional
circuit $FA_3$. For $k=1$ we need a circuit $SFA_3$ instead of
$FA_3$, since summation in the previous column involves $MDFA$.

The use of $MDFA$ requires a conversion to the special bit
encoding $(x,y) \to (x,x\oplus y)$. All $MDFA$ outputs encoded
this way may be connected to $MDFA$ inputs of the same encoding,
with the exception of the last $MDFA$, in the $(2n-2)$-th column.
Therefore, to execute all summations we need additionally $q+1$
$XOR$ gates, where $q$ is a number of $MDFA$.

Therefore, if $n\ge 4$, then the second stage of multiplication
utilizes $n$ circuits $HA$, $n-3+2(n\bmod2)$ circuits $FA_3$, one
circuit $SFA_3$, $q=(n^2-3n)/2+1-(n\bmod2)$ circuits $MDFA$ and
$q+1$ $XOR$ gates (the number of $MDFA$ is easy to derive from the
number of $(3,2)$-CSA, since $MDFA$ reduces the total number of
summand bits by 2, $FA_3$ or $SFA_3$ reduces it by 1; the number
of summand bits before summation stage is $n^2$, and at the end it
is $2n$). Summing up the complexities of subcircuits we can bound
the complexity $M(n)$ of the multiplication circuit as
$$ M(n) \le 5,5n^2-6,5n-1+(n\bmod 2). $$
The estimate holds also for $2\le n\le3$.

{\bf Karatsuba method.} Represent two $m$-bit multiplication
operands as $A_12^n+B_1$ and $A_22^n+B_2$, where $n=\lceil m/2
\rceil$, $0 \le B_i < 2^n$, $0 \le A_i < 2^{m-n}$. Then, the
product may be computed by the formula:
$$ A_1A_22^{2n} + ((A_1+B_1)(A_2+B_2)-A_1A_2-B_1B_2)2^n + B_1B_2. $$

The implied circuit consists of two addition circuits computing
$A_1+B_1$ and $A_2+B_2$, three multiplication subcircuits for
$(n+1)$-bit, $(m-n)$-bit and $n$-bit operands, and a subcircuit
for the final addition-subtraction. The structure of this final
addition is shown on the pattern below (see Fig. 2). Symbols
``$+$'' and ``$-$'' denote summand and subtrahend bits,
respectively. Columns are indexed so that an index $i$ corresponds
to a bit with weight $2^i$. Pairs of bits in brackets are missing
when $m$ is odd.

\[
\begin{matrix}
(++)+\cdots  + +  +  + \: + \, \cdots + \phantom{+ \cdots +}                                      & A_1A_22^{2n} \\
\phantom{(++)+\cdots}       + +  +  + \: + \, \cdots +  + \, \cdots +      & (A_1+B_1)(A_2+B_2)2^n  \\
\phantom{(++)+\cdots  + +}  \:   (- -) -  \cdots  -  -   \cdots -                                   & -A_1A_22^n \\
\phantom{(++)+\cdots  + +(}       -  - - \, \cdots  -  - \cdots  -                                      & -B_1B_22^n \\
\phantom{(++)+\cdots  + +  +  + \: + \, \cdots + ++}  \;\; \,   +   \cdots +     \cdots +                                          &  B_1B_2 \\
\scriptstyle \qquad \; 4n-1  \qquad  \qquad  \quad  3n-1  \qquad \qquad   2n-1 \quad \;\;\, n \qquad  0                                       &  \\
\end{matrix}
\]
\[ \text{Fig. 2. Pattern of the final addition-subtraction in the Karatsuba method}  \]

Consider the following column summation rule for the final
addition: if there exist 3 summand bits and 2 subtrahend bits, use
$MDFA^-$; else, if there exist 3 bits, use a suitable circuit from
the $FA_3$ family; else, if there exist 2 bits, use an appropriate
circuit from the $HA$ family.

Denote by $h^+(k)$, $h^-(k)$ numbers of summand and subtrahend
bits in the $k$-th column after completion of summation in all
lower-order columns. One can easily verify that $h^+(n)=h^-(n)=2$,
$h^+(n+1)=h^-(n+1)=3$ and $h^+(k)=3$, $h^-(k)=4$ for $n+2\le k \le
2m-n-1$. We use one $FA_3^-$ and one $HA$ in the $n$-th column,
one $MDFA^-$ and one $HA^{\pm}$ in the $(n+1)$-th column, one
$MDFA^-$ and one $FA_3^-$ in any subsequent column up to
$(2m-n-1)$-th.

When $m$ is odd, we have $h^+(k)=h^-(k)=3$ for $k=3n-1,\,3n-2$.
Therefore, one $MDFA^-$ and one $HA^{\pm}$ should be used in the
corresponding columns.

Further, $h^+(3n)=3$, $h^-(3n)=2$ (use $MDFA^-$), $h^+(3n+1)=3$,
$h^-(3n+1)=1$ (use $SFA_3^-$ and $HA^{\pm}$), $h^+(3n+2)=2$,
$h^-(3n+2)=1$ (use $FA_3^0$, since the value
$(A_1+B_1)(A_2+B_2)-A_1A_2-B_1B_2$ is non-negative). At last,
$h^+(k)=2$, $h^-(k)=0$ for $3n+3\le k \le 2m-1$: use $HA$
elsewhere, but use $XOR$ in the most significant column, since no
carry is required there.

As in the standard method, conversion to the special pair-of-bits
encoding requires $q+1$ additional $XOR$ gates, where $q$ is the
number of $MDFA^-$ subcircuits.

Now, we're going to estimate the complexity $K(m)$ of the
multiplication circuit. It's a common knowledge, that the
complexity of the addition of two $n$-bit numbers is $5n-3$, the
addition of an $n$-bit number and an $(n-1)$-bit number has
complexity $5n-6$ (see e.g.~\cite{ga}). If $m\ge 10$, then the
final Karatsuba adder-subtractor contains $n-1$ circuits $HA$ or
$HA^{\pm}$, $2m-2n-1$ circuits $FA_3^-$, one circuit $SFA_3^-$,
one circuit $FA_3^0$, $2n$ circuits $MDFA^-$ and $2n+2$ $XOR$
gates.

Note, that one can save some gates. Columns indexed by $n+i$ and
$2n+i$, for $i=0,\ldots,n-1$, contain identical pairs of bits
(from summands $B_1B_2$, $-A_1A_22^n$ and $-B_1B_22^n$,
$A_1A_22^{2n}$, respectively). Arrange the computation process to
pass these bits to the inputs of $FA_3^-$ in columns indexed by
$n$ and $n+2,\ldots,2m-n-1$, and to the inputs of $MDFA^-$ encoded
by $(x,\,x\oplus y)$ in other columns (that is, in $(n+1)$-th
column, and in the case of odd $m$, also in $(3n-2)$-th and
$(3n-1)$-th columns).

Thus, for $i=0$ and $2\le i \le n-1-2(m\bmod2)$ we can save two
gates $XOR$ and $ANDNOT$ via exploiting a CSA from Fig. 3a in a
lower-order column and a CSA from Fig. 3b in a higher-order column
(CSA's are different since the signs of bits in low-order and
high-order columns are rearranged). For $i=1$ and $n-2(m\bmod2)\le
i \le n-1$, one $XOR$ gate is to be saved.

\begin{figure}[htb]
\begin{picture}(350,120)(0,0)

\put(70,20){\begin{picture}(100,100)
\multiput(50,75)(25,0){2}{\circle{10}}
\multiput(25,50)(25,0){2}{\circle{10}} \put(75,25){\circle{10}}
\put(25,30){\line(0,1){15}}
\multiput(25,55)(25,25){2}{\line(0,1){15}}
\multiput(29,54)(25,25){2}{\line(1,1){17}}
\put(75,84){\line(0,1){11}} \put(75,82){\circle{4}}
\put(50,59){\line(0,1){11}} \put(50,57){\circle{4}}
\put(71.2,28.8){\line(-1,1){17.4}}
\put(46.5,53.5){\line(-1,1){17}}
\put(71.2,78.8){\line(-1,1){17.4}} \put(75,70){\line(0,-1){40}}
\put(75,20){\line(0,-1){15}}
\multiput(20,50)(25,25){2}{\line(1,0){10}}
\put(70,25){\line(1,0){10}} \put(75,20){\line(0,1){10}}
\multiput(25,45)(25,25){2}{\line(0,1){10}}
\multiput(46.3,47.5)(25,25){2}{$\wedge$} \put(20,76){$x_2$}
\put(45,100){$x_1$} \put(72,99){$x_3$} \put(22,23){$v$}
\put(72,-2){$u$} \end{picture}}

\put(102,0) {Fig. 3a}

\put(220,20){\begin{picture}(100,100)
\multiput(50,75)(25,0){2}{\circle{10}}
\multiput(25,50)(25,0){2}{\circle{10}} \put(75,25){\circle{10}}
\multiput(25,30)(25,25){2}{\line(0,1){15}}
\multiput(25,55)(25,25){2}{\line(0,1){15}}
\multiput(29,54)(25,25){2}{\line(1,1){17}}
\put(75,84){\line(0,1){11}} \put(75,82){\circle{4}}
\put(71.2,78.8){\line(-1,1){17.4}}
\put(71.2,28.8){\line(-1,1){17.4}}
\put(46.5,53.5){\line(-1,1){17}} \put(75,70){\line(0,-1){40}}
\put(75,20){\line(0,-1){15}}
\multiput(20,50)(25,25){2}{\line(1,0){10}}
\put(70,25){\line(1,0){10}} \put(75,20){\line(0,1){10}}
\multiput(25,45)(25,25){2}{\line(0,1){10}} \put(46.3,46.1){$\vee$}
\put(71.3,72.5){$\wedge$} \put(20,76){$x_2$} \put(45,100){$x_3$}
\put(72,99){$x_1$} \put(22,23){$v$} \put(72,-2){$u$}
\end{picture}}

\put(252,0) {Fig. 3b}

\end{picture}
\end{figure}

So, the following recurrent formulae hold:
\begin{align}\label{rec}
K(2n-1) &\le K(n+1) + K(n) + K(n-1) + 38n - 16, \qquad &n\ge6, \notag \\
K(2n) &\le K(n+1) + 2K(n) + 38n - 2, \qquad &n\ge5.
\end{align}

A halving iteration of the Karatsuba method provides an advantage
when $m=16$ or $m\ge18$. Bounds on the complexity $L(m)$ of
multiplication of $m$-bit numbers for $m\le 18$ are collected in
the Table 1 (symbol $^*$ indicates values obtained via Karatsuba
method).

\begin{table}[htb]
\caption{Bounds for the complexity of multiplication of $m$-bit
numbers}
\begin{tabular}{||c||c|c|c|c|c|c|c|c|c||}
 \hline
  $m$ & 1 & 2 & 3 & 4 & 5 & 6 & 7 & 8 & 9 \\
 \hline
  $L(m)$ & 1 & 8 & 30 & 61 & 105 & 158 & 224 & 299 & 387 \\
 \hline \hline
  $m$ & 10 & 11 & 12 & 13 & 14 & 15 & 16 & 17 & 18 \\
 \hline
  $L(m)$ & 484 & 594 & 713 & 845 & 986 & 1140 & $1287^*$ & 1479 & $1598^*$ \\
 \hline\end{tabular}
\end{table}

For the convenience of comparison, let us derive the complexity of
the Karatsuba multiplication circuit in an explicit form for
$m=2^k$. Denote
$$ X_k = \begin{bmatrix} K(2^k+2) \\ K(2^k+1) \\ K(2^k) \end{bmatrix}, \qquad
 A = \begin{bmatrix} 1 & 2 & 0 \\ 1 & 1 & 1  \\ 0 & 1 & 2 \end{bmatrix}, \qquad
  b_k = \begin{bmatrix} 38\cdot2^k+36 \\ 38\cdot2^k+22 \\ 38\cdot2^k-2 \end{bmatrix}. $$
Recurrences (\ref{rec}) imply $X_{k+1} \le AX_k+b_k$ for $k\ge4$.
Via common calculations, we obtain (\ref{kar}) as the solution of
the latter inequality (initial values of the complexity should be
taken from the Table 1).

Research supported in part by RFBR, grant 14--01--00671a.


\begin{thebibliography}{99}

\bibitem{ka}
A.A. Karatsuba, Yu.P. Ofman. Multiplication of multidigit numbers
on automata. {\it DAN USSR} {\bf145}(2) (1962), 293--294 (in
Russian). Eng. transl. in {\it Soviet Phys. Dokl.} {\bf 7} (1963),
595--596.

\bibitem{ga}
S.B. Gashkov. Entertaining computer arithmetic. Fast algorithms
for operations with numbers and polynomials. Moscow, Librocom,
2012 (in Russian).

\bibitem{dkky}
E. Demenkov, A. Kojevnikov, A. Kulikov, G. Yaroslavtsev. New upper
bounds on the Boolean circuit complexity of symmetric functions.
{\it Inf. Proc. Letters} {\bf 110}(7) (2010), 264--267.

\end{thebibliography}
\end{document}